\documentclass[11pt]{article}
\usepackage[dvips]{color}
\usepackage{epsfig}
\usepackage{amsmath}
\usepackage{graphicx}
\usepackage{amsfonts}
\date{}

\begin{document}
\title{{\bf Polymeric Quantization and Black Hole Thermodynamics}}
\author{M. A. Gorji$^1$\thanks{%
e-mail: m.gorji@stu.umz.ac.ir},\,\, K. Nozari$^1$\thanks{%
e-mail: knozari@umz.ac.ir}\,\, and B. Vakili$^2$\thanks{%
e-mail: b-vakili@iauc.ac.ir}
\\\\
$^1${\small {\it Department of Physics, Faculty of Basic Sciences,
University of Mazandaran,}}\\
{\small {\it P.O. Box 47416-95447, Babolsar, Iran}}\\
$^2${\small {\it Department of Physics, Chalous Branch, IAU, P.O.
Box 46615-397, Chalous, Iran}}}\maketitle

\begin{abstract}
Polymer quantization is a non-standard representation of the quantum
mechanics that inspired by loop quantum gravity. To study the
associated statistical mechanics, one needs to find microstates'
energies which are eigenvalues of the Hamiltonian operator in the
polymer framework. But, this is not an easy task at all since the
Hamiltonian takes a nonlinear form in polymer picture. In this
paper, we introduce a semiclassical method in which it is not
necessary to solve the eigenvalue problem. Instead, we work with the
classical Hamiltonian function and the deformed density of states in
the polymeric phase space. Implementing this method, we obtain the
canonical partition function for the polymerized systems and we show
that our results are in good agreement with those arising from full
quantum considerations. Using the partition function, we study the
thermodynamics of quantum Schwarzschild black hole and we obtain
corrections to the Bekenstein-Hawking entropy due to loop quantum
gravity effects.
\vspace{5mm}\noindent\\
PACS numbers: 04.60.Pp,\, 04.60.Bc,\, 04.70.Dy\vspace{0.8mm}
\end{abstract}

\section{Introduction}
Our current understanding of gravity is based on the Einstein
general theory of relativity. The essential feature of General
Relativity (GR) is that gravity forms the geometry and geometry
tells the particles how to move in spacetime. In this sense, while
GR seems to be a purely classical theory, in its most important
applications, say, cosmology and Black Hole (BH) theories, the
system under consideration essentially obeys the quantum mechanical
rules. Quantization of gravity is therefore the main challenge of
theoretical physics community. In the absence of a full theory of
quantum gravity (QG), however, there have been a variety of
approaches to the issue; among them is the canonical quantum theory
of gravity first introduced by De Witt \cite{CQC}. Also, string
theory, Loop Quantum Gravity (LQG) and noncommutative geometry, as
other approaches to QG proposal, have revealed some unknown aspects
of ultimate QG scenario \cite{QGML,LQG}. One common feature of all
approaches to QG proposal is the existence of a minimal measurable
length (preferably of the order of the Planck length) \cite{LQG2}.
In recent years, some phenomenological aspects of effective QG
candidates have been introduced in quantum mechanics through
deformation of algebraic structure of ordinary quantum mechanics.
For instance the generalized uncertainty principle \cite{GUP1}, and
noncommutative geometry \cite{NCG1} are the most well-known
deformations that impose the ultraviolet and infrared cutoffs for
the physical systems \cite{UV1}. Among the models that deal with the
idea of the existence of a minimal measurable length scale, the
so-called polymer quantization of a dynamical system uses a method
much similar to the effective models of LQG
\cite{Ashtekar,Corichi1}. The minimal length scale, here known as
the polymer length scale, is encoded in the Hamiltonian of the
system and thus instead of a deformed algebraic structure coming
from the noncommutative phase space variables, deformation shows
itself in the Hamiltonian function. With this motivations, polymer
quantization has attracted some attentions in recent years in the
fields dealing with the quantum gravitational effects in a physical
system. Specially, in the spirit of quantum features of cosmological
models, the polymeric effects provide a modified Friedmann equation
which is very similar to the one coming from loop quantization of
the model \cite{Corichi1,Campiglia,PLCosmology}. Also, through the
lines sketched by the polymer quantum mechanics, it is expected that
applying the polymer quantization approach to the quantum
gravitational systems, including thermodynamics of BHs and other
statistical systems, leads one to some results which may be
comparable with those obtained from more fundamental LQG and string
theories (see \cite{PLBH,PLBHEntropy,PLSTAT,Acosta,PLuncertainty,
PLGUP} for some investigations of BH thermodynamics in the polymer
picture).

Indeed, formulation of BH thermodynamics is based on quantum field
theoretical considerations in the curved spacetime leading to the
well-known result of BH thermal emission through Hawking process
\cite{Hawking}. The discovery of BH evaporation makes this system to
be a thermodynamical system and consequently one can extract the
thermodynamical quantities from the statistical mechanics
considerations. This issue was first studied by Bekenstein that
showed the relation between the horizon area of BH and the number of
accessible microstates \cite{Bekenstein}. To implement the standard
statistical mechanics methods, one needs the Hamiltonian formulation
of the BH that is not generally formulated even in the classical
framework of GR. Nevertheless, it is possible to define Hamiltonian
for the some particular cases, such as Schwarzschild BH
\cite{Makela,Louko}. Through this procedure, quantization of the
Schwarzschild mass is realized which coincides with Bekenstein
proposal \cite{Bekenstein}. Now, polymer quantization provides a new
framework to extract all these results satisfactorily with the
advantage that it contains quantum gravity effects from very
beginning. Nevertheless, some technical difficulties arise due to
the nonlinear form of the polymeric Hamiltonian
\cite{Corichi1,Husain}. In this paper, we formulate a semiclassical
statistical mechanics in polymer framework and we show that our
results coincide with those obtained from full quantum
considerations in the limit of high temperature. This method is
applicable to study the thermodynamics of the quantum BHs. The
advantage of this semiclassical procedure is that it is not
necessary to solve the Hamiltonian eigenvalue problem. Implementing
this method, we study the thermodynamics of the quantum
Schwarzschild BH in polymer framework and we find corrections to the
Hawking temperature and entropy.

\section{Polymer representation of quantum mechanics}

Almost all candidates for QG predict that the spacetime, when it is
considered in the high energy limit, has a discrete structure. This
is due to the existence of natural cutoffs such as the minimal
measurable length. In ordinary Schr\"{o}dinger representation of
quantum mechanics, the position and momentum operators have
continuous spectrum and therefore one can work in position or
momentum space representations. However, when one considers the
quantum gravitational effects, these representations are no longer
applicable. For instance, position space representation fails to be
applicable due to fuzzy nature of spacetime manifold in the presence
of a minimal measurable length \cite{fuzzy}. In other words,
existence of a minimum measurable length of the order of Planck
length prevents complete resolution of spacetime points, leading
naturally to generalization of the Heisenberg uncertainty principle
\cite{GUP1,UV1}.

Polymer quantization provides an alternative framework to study the
mathematical and physical aspects of the discrete spaces
\cite{Ashtekar,Corichi1}. In what follows we present a brief review
of this issue and its semiclassical outcomes.

Suppose an abstract ket $|\nu\rangle$, labeled by a real number
$\nu$, which belongs to the nonseparable Hilbert space
${\mathcal{H}}_{\rm poly}$. Considering a finite collections of
vectors $|\nu_i\rangle$ where $\nu_i\in\,{\mathbb R}$ and
$i=1,2,...,N$, one can construct an appropriate state by taking
a linear combination of these vectors as
\begin{equation}\label{state}
|\psi\rangle=\,\sum_{i=1}^{N}a_i\,|\nu_i\rangle\,.
\end{equation}
After introduction of the abstract Hilbert space of the model, we
can consider the wave function in this framework. Suppose a system
with a phase space with coordinates $x$ and conjugate momenta $p$.
In momentum representation, states will be denoted by
$\psi(p)=\langle\,p|\psi\rangle$ in a nonseparable Hilbert space
${\mathcal{H}}_{\rm poly}$ as
\begin{eqnarray}
\psi_{\nu}(p)=\langle\,p|\nu\rangle=e^{i\nu p/\hbar}.\nonumber
\end{eqnarray}
One can define one-parameter family of exponential shift operator
$\hat{V}=e^{i\lambda p/\hbar}$ in the Hilbert space ${\mathcal{H}}_{
\rm poly}$ which acts on the states as
\begin{equation}\label{exp-op}
\hat{V}(\lambda).\,\psi_{\nu}(p)=e^{i\lambda p/\hbar}\,
e^{i\nu p/\hbar}=e^{i(\nu+\lambda)p/\hbar}=\psi_{\nu+\lambda}(p)\,.
\end{equation}
The next step is to consider the dynamics of the system in the
polymer representation which needs the Hamiltonian of the
corresponding underlying quantum theory. We start with the classical
Hamiltonian
\begin{equation}\label{nhamiltonian}
H=\frac{p^2}{2m}+U(x)\,,
\end{equation}
which is a smooth function of the phase space variables $x$ and $p$.
The standard method of quantization is to replace the classical phase
space variables with corresponding operators. In polymer
representation, however, this argument fails to be applicable,
because the corresponding momentum operator does not exist. The main
task now is to define momentum operator and the square of the
momentum in polymer framework. The standard prescription is to define
the lattice $\gamma_{\mu_0}$ on the configuration space as $\gamma_{
\mu_0}=\,\{x\in {\mathbb{R}}\,|\,x=\,n\mu_0,\,\forall\,n\in
{\mathbb{Z}}\}$, which ensures the discreteness of the position $x$
with discreteness parameter $\mu_0$. The states in Hilbert space
${\mathcal{H}}_{\gamma_{\mu_0}}$ are given by\footnote{
${\mathcal{H}}_{\gamma_{\mu_0}}$ is a separable subspace of the
larger nonseparable Hilbert space ${\mathcal{H}}_{\rm poly}$.}
\cite{Corichi1}
\begin{equation}\label{PLS}
|\psi\rangle=\,\sum_{n} b_n |\mu_n\rangle\,,
\end{equation}
where coefficients $b_n$ satisfy $\sum_n|b_n|^2<\infty$. Since in QG
scales the space becomes discrete, a natural question then arises:
Is it possible to recover the ordinary Schr\"{o}dinger
representation in low energy regime? This problem, known as the
continuum limit of the theory, is studied in
\cite{Corichi1,Campiglia, Corichi} and it is shown that the theory
should be formulated in such a way that the standard Schr\"{o}dinger
representation is recovered in the limit of small values of the
discreteness parameter $\mu_0 \rightarrow\,0$. Furthermore, note
that since momentum is the generator of translation we should have a
momentum operator that generates discrete displacement according to
the relation (\ref{PLS}), such that the results remain in the
lattice $\gamma_{\mu_0}$. How can we apply this setup to form the
momentum operator? This may be done by the exponential shift
operator $\hat{V}(\mu_0)$ as
\begin{equation}\label{PLSO}
\hat{V}(\mu_0)\,|\mu_n\rangle =e^{i\mu_0 p/\hbar}e^{i\mu_n
p/\hbar}=e^{i(\mu_0+\mu_n)p/\hbar}=\,|\mu_n+\mu_0\rangle=
|\mu_{n+1}\rangle.
\end{equation}
So, the momentum operator can be defined by means of the shift
operator as \cite{Ashtekar},
\begin{equation}\label{PLMO}
\hat{p}_{\mu_0}\,|\mu_n\rangle=\,\frac{\hbar}{2i{\mu_0}}
\left[\hat{V}(\mu_0)-\hat{V}(-\mu_0)\right]|\mu_n\rangle
=\frac{i\hbar}{2\mu_0}\Big(| \mu_{n+1}\rangle-|\mu_{n-1}
\rangle\Big)\,.
\end{equation}
To consider the kinetic term in the Hamiltonian
(\ref{nhamiltonian}), one needs the squared momentum which can be
obtained as \cite{Ashtekar}
\begin{equation}\label{PLMO2}
\hat{p}_{\mu_0}^2\,|\mu_n\rangle=\,\frac{\hbar^2}{
\mu_0^2}\left[2-\hat{V}(\mu_0)-\hat{V}(-\mu_0)\right]
|\mu_n\rangle=\frac{\hbar^2}{\mu_0^2}\Big(2|\mu_{
n}\rangle-|\mu_{n+1}\rangle-|\mu_{n-1}\rangle\Big).
\end{equation}
The quantum polymeric Hamiltonian operator can be deduced from the
classical Hamiltonian (\ref{nhamiltonian}) through the relation
(\ref{PLMO2}) as
\begin{equation}\label{EV-QH}
\hat{H}_{\mu_0}=\frac{\hbar^2}{2m\mu_0^2}\left[2-
\hat{V}(\mu_0)-\hat{V}(-\mu_0)\right]+\hat{U}(x)\,.
\end{equation}
Now one can replace the shift operator in the relations (\ref{PLMO})
and (\ref{PLMO2}) with the exponential function through the relation
(\ref{exp-op}) as $\hat{V}(\mu_0)\rightarrow\, e^{i\mu_0 p/\hbar}$,
in order to get an approximation for the momentum and its square
\cite{Corichi1,Campiglia}
\begin{equation}\label{momentum1}
\hat{p}\,\longrightarrow\,\frac{e^{\widehat{i\mu p}}
-e^{\widehat{-i\mu p}}}{2i\mu}=\frac{\widehat{\sin
(\mu\,p)}}{\mu},
\end{equation}
\begin{equation}\label{momentum2}
\hat{p}^2\,\longrightarrow\,\frac{1}{\mu^2} \Big(2\,-e^{\widehat{i\mu
p}}-e^{\widehat{-i\mu p}}\Big) =\frac{2}{\mu^2}\left[1-\widehat{\cos(\mu
p)}\right],
\end{equation}
where $\mu=\mu_0/\hbar$ and the approximations are reliable for $\mu
p\ll{1}$  to recover the standard continuum theory in the limit of
small discreteness parameter \cite{Corichi1}. By substituting from
the relation (\ref{momentum2}) into the relation (\ref{EV-QH}), the
polymeric quantum Hamiltonian can be approximated as
\begin{equation}\label{qhamiltonian}
H_{\mu}=\,\frac{1}{m\mu^2}\left[1\,-\,\widehat{\cos(\mu p)}\right]
+\hat{U}(x).
\end{equation}
The replacements (\ref{momentum1}) and (\ref{momentum2}) suggest the
idea that a classical polymeric theory may be obtained via the
process that is dubbed usually as \emph{Polymerization} in
literature \cite{Corichi2}
\begin{equation}\label{polymerization}
{\mathcal P}[x]=x,\hspace{5mm}{\mathcal P}[p]=\frac{\sin(\mu p)}{
\mu},\hspace{5mm}{\mathcal P}[p^2]=\frac{2}{\mu^2}\left[
1-\cos(\mu p)\right],
\end{equation}
where now $(x,p)$ are the classical phase space variables. Clearly,
the momenta are bounded as $-\frac{\pi}{\mu}<p<\frac{ \pi}{\mu}$ in
the polymer framework. Applying the transformations
(\ref{polymerization}) to the classical Hamiltonian
(\ref{nhamiltonian}) gives
\begin{equation}\label{hamiltonian}
H_{\mu}=\,\frac{1}{m\mu^2}\left[1\,-\,\cos(\mu p)\right]+U(x),
\end{equation}
which coincides with Hamiltonian (\ref{qhamiltonian}) in the
semiclassical regime. So, a one-parameter $\mu$-dependent classical
theory can be obtained from the polymerization process without any
direct reference to the quantum ones. Indeed, in this approach
instead of first finding the quantum solutions in the polymer
quantization framework based on the Hamiltonian operator
(\ref{EV-QH}), one modifies the classical Hamiltonian
(\ref{nhamiltonian}) according to the polymerization process
(\ref{polymerization}) to get an effective polymer-deformed
Hamiltonian (\ref{hamiltonian}) and then deals with classical
dynamics of the system with this deformed Hamiltonian. In the
corresponding classical system the discreteness parameter $\mu$
plays an important role since its existence supports the idea that
the $\mu$-correction to the classical theory is a signal from QG. In
this sense, we emphasize that the classical Hamiltonian and also the
classical equations of motion should be recovered in the continuum
limit $\mu\rightarrow\,0$. In summary, to study the quantum effects
in a given classical system there is a simple strategy: one
constructs a suitable one parameter family of an effective
Hamiltonian $H_{\mu}$ with correct continuum limit based on which
the resulting classical theory shows the quantum effects, with no
direct reference to the corresponding quantum theory. This strategy
seems to work well specially on some cosmological models that are
treated, for example in \cite{PLGUP} and \cite{Corichi2}, and the
resulting equations of motion are in good agreement with the one
coming from the quantization of the system by LQG methods. In the
following, we will use this strategy to study the thermodynamics of
the physical systems in polymer framework.

\section{Polymeric statistical mechanics}

Quantum mechanics determines the microstates of the physical
systems, and all thermodynamical quantities attributed to a given
system can be extracted from its statistical properties through the
quantum partition function
\begin{equation}\label{QPF}
Z=\sum_{\varepsilon}\,e^{-\beta\,\varepsilon},
\end{equation}
where $\varepsilon$ are the energy eigenvalues of the microstates.
In our case, these eigenvalues of the polymer-deformed Hamiltonian
operator (\ref{EV-QH}) are given by the relation
\begin{equation}\label{EVP}
\hat{H}_{\mu_0}|\psi\rangle=\varepsilon|\psi\rangle\,,
\end{equation}
where the corresponding physical states are defined by the relation
(\ref{PLS}) on the polymeric Hilbert space ${\mathcal H}_{\rm poly}$.
In fact, solving the eigenvalue problem (\ref{EVP}) is not an easy
task at all (see for example \cite{Corichi1,Husain}). Nevertheless,
inspired by the idea of polymerization we can seek for an
alternative approach to consider the quantum effects, that is, to
work with the classical Hamiltonian (\ref{hamiltonian}) and
associated density of states in the semiclassical regime. The
advantage of this method is that it does not need to solve the
eigenvalue problem (\ref{EVP}) in the polymer picture. Furthermore,
we introduce the transformation in the polymeric phase space which,
as we will see, is more understandable from the statistical point
of view.

Consider the noncanonical transformation
\begin{eqnarray}\label{transformation}
\big(x\,,\,p\big)\,\longrightarrow\,\Big(X=x\,
\,,\,P=\frac{2}{\mu}\sin\big(\frac{\mu p}{2}\big)\Big)\,,
\end{eqnarray}
in the polymeric phase space with Hamiltonian (\ref{hamiltonian}).
In the transformation (\ref{transformation}), $(x,p)$ are the usual
canonical phase space variables and $(X,P)$ are the new noncanonical
variables that we call them polymer variables. From the
transformation (\ref{transformation}), it is clear that the polymer
momentum should be bounded as $-\frac{2}{\mu}<P<\frac{ 2}{\mu}$,
since the canonical momentum is bounded as $-\frac{\pi}{
\mu}<p<\frac{\pi}{\mu}$ in polymer phase space. More precisely, the
topology of the momentum part of the polymer phase space is $S^1$
rather than usual $\mathbb{R}$ and there is always a maximal
momentum for the system under consideration in this framework
\cite{Corichi1}. This result of the polymeric systems is comparable
to the predictions of other theories such as the ones studied in
\cite{Batt08} and \cite{Doubly}. The reason for introducing the
transformation (\ref{transformation}) becomes clear if one notes
that the effective Hamiltonian (\ref{hamiltonian}) gets the standard
form
\begin{equation}\label{NDhamiltonian}
H_{\mu}=\,\frac{P^2}{2m}+U(X),
\end{equation}
where one adopts the polymer variables $(X,P)$ and consequently all
the polymeric effects are summarized in the density of states. To
obtain the corresponding density of states, one needs to the
Jacobian of the transformation (\ref{transformation}) which is given
by \cite{CCC2}
\begin{equation}\label{Jacobian}
J=\,\frac{\,\partial(X\,,P)}{\partial(x\,,p)}=\,\cos(\mu p/2)=\sqrt{
1-(\mu P/2)^2}\,.
\end{equation}
Having the transformation's Jacobian, the polymer-deformed density
of states can be obtained as
\begin{equation}\label{dos}
\frac{1}{h}\int_{|p|<\frac{\pi}{\mu}}dx\,dp\,\longrightarrow\,\frac{
1}{h}\int\frac{dX\,dP}{J}=\,\frac{1}{h}\int_{|P|<\frac{2}{\mu}}\frac{
dX\,dP}{\sqrt{1-(\mu P/2)^2}},
\end{equation}
where $h=2\pi\hbar$ is the Planck constant. As we said before, from
the relations (\ref{NDhamiltonian}) and (\ref{dos}) it is clear that
all the polymeric effects are summarized in the density of states
when one implements the polymer variables $(X,P)$. The density of
states determines the number of microstates for the statistical
systems and consequently the polymeric effects change only the
number of accessible microstates for the statistical system. Also,
it is important to note that the measure (\ref{dos}) is nonsingular
since the momentum is bounded as $|P|<\frac{2}{\mu}$. From another
perspective, this can be viewed as a redefinition of the Planck
constant as an effective momentum dependent quantity ${\hbar}^{\rm
(poly)}=\hbar\sqrt{ 1-(\mu P/2)^2}$. This result is in some sense
similar to the results coming from the generalized uncertainty
principle and noncommutative phase space as have been obtained in
\cite{Adler1}.

Now one can find immediately the polymeric partition function from
the polymer-modified states density (\ref{dos}) and the associated
Hamiltonian function (\ref{NDhamiltonian}) as
\begin{equation}\label{pf}
Z^{\rm (poly)}=\frac{1}{h}\int\exp[-\beta U(X)]\,dX\times\int_{-\frac{2}{
\mu}}^{-\frac{2}{\mu}}\frac{\exp\big[-\frac{\beta P^2}{2m}\big]\,dP}{
\sqrt{1-(\mu P/2)^2}}\,.
\end{equation}
For the one-dimensional harmonic oscillator with the potential
$U(X)=\frac{ 1}{2}m\omega^2X^2$, where $m$ and $\omega$ denote the
mass and frequency respectively, the corresponding partition
function can be obtained from the relation (\ref{pf}) as
\begin{equation}\label{pfolho0}
Z^{\rm (poly)}=\frac{1}{\hbar\omega\mu}\Big(\frac{2\pi}{m\beta}\Big)^{1/2}
\,I_0\Big[\frac{\beta}{m\mu^2}\Big]\exp\Big[-\frac{\beta}{m\mu^2}\Big],
\end{equation}
where $I_0$ denotes the first kind of the modified Bessel functions.
In the limit of the small discreteness parameter,
$\mu\rightarrow\,0$, the above partition function gives
\begin{eqnarray}\label{pfolho}
Z^{\rm (poly)}=\,(\beta\hbar\omega)^{-1}\,\Big(\,1
\,+\,\frac{\mu^2}{8}\Big(\frac{m}{\beta}\Big)\,+\,{
\mathcal{O}}[\mu^4]\,\Big),
\end{eqnarray}
where the first term is the usual semiclassical partition function for the
one-dimensional harmonic oscillator and the second term is the polymeric
correction.

At this step, it is interesting to compare the result (\ref{pfolho})
with the corresponding result coming from full quantum
considerations. In quantum picture, one has to solve the eigenvalue
problem (\ref{EVP}) when the system under study is a one-dimensional
harmonic oscillator. The energy eigenvalues for the one-dimensional
harmonic oscillator in the limit of $\mu\rightarrow\,0$ is obtained
in the Ref. \cite{Husain} that are given by
\begin{equation}\label{SHO-EV}
\varepsilon_n=\Big(n+\frac{1}{2}\Big)\hbar\omega-\Big(\frac{2n^2+2n+1
}{32}\Big)m\mu^2\hbar^2\omega^2\,.
\end{equation}
Substituting the above result into relation (\ref{QPF}) gives the quantum
partition function of the polymeric one dimensional harmonic oscillator
as \cite{Acosta}
\begin{equation}\label{QPF2}
Z^{\rm (poly)}=\frac{e^{-\beta\hbar\omega/2}}{1-e^{-\beta\hbar
\omega}}\,\Bigg(\,1+\,\frac{1}{32}m\mu^2\hbar^2\omega^2\beta
\bigg(\frac{1+e^{-\beta\hbar\omega}}{1-e^{-\beta\hbar\omega}}
\bigg)^2\,+\,{\mathcal{O}}[\mu^4]\Bigg).
\end{equation}
In the limit of high temperature, $\beta\rightarrow\,0$, where the
semiclassical and quantum statistics are going to be equivalent,
relation (\ref{QPF2}) exactly coincides with the relation
(\ref{pfolho}). It is important to note that, one should doing some
tedious calculation to obtain the result (\ref{SHO-EV}) (see
\cite{Husain}) and then performing the summation over energy
eigenvalues (\ref{SHO-EV}) in quantum partition function (\ref{QPF})
to obtain the result (\ref{QPF2}) which is not an easy task at all
(see \cite{Acosta}). However, the method we have introduced gives
the same results but now in a more simpler manner. While the full
quantum results preserve their importance for the other goals
\cite{Ashtekar, Corichi1,Husain}, our method is more suitable when
one concerns for the statistical considerations of the polymeric
systems. In the next section, we shall see that the quantum
Schwarzschild BH can be modeled with a one-dimensional harmonic
oscillator and hence, the relation (\ref{pfolho}) may be used to
study its thermodynamical properties in the polymer quantization
scheme.

\section{Thermodynamics of quantum black holes}
\subsection{Quantization of the Schwarzschild black hole}

As a canonical system, a Schwarzschild BH can be described by a
single canonical pair $(m,p_{m})$, where $m$ may be identified with
its mass. In Ref. \cite{Makela} it is shown, under some conditions,
that the numerical value of the Hamiltonian of such a system is
equal to its mass, that is, $H=m$. However, there is a canonical
transformation (again introduced in \cite{Makela}) from $(m,p_{m})$
to a new canonical pair $(a,p_a)$ such that
\begin{equation}\label{revq1}
|p_m|=\sqrt{2ma-a^2}+m\arcsin\left(1-\frac{a}{m}\right)+\frac{
\pi}{2}m,\hspace{5mm}p_a=sgn(p_m)\sqrt{2ma-a^2},
\end{equation}which in terms of the new variables the classical
Hamiltonian takes the following form
\begin{equation}\label{revq2}
H=\frac{p_{_a}^2}{2a}+\frac{1}{2}a.
\end{equation}
Therefore, according to the canonical quantization formalism, the
Wheeler-De Witt (WDW) equation for the Schwarzschild BH can be
written in the form $\hat{H}\Psi=m\Psi$, which in ordinary units
reads
\begin{equation}\label{wd1}
\frac{\hbar^2 G^2}{c^6} a^{-s-1} \frac{d}{da}\Big(a^s\frac{d}{da}
\Psi(a)\Big)\,=\,\Big(a-\frac{2GM}{c^2}\Big) \Psi(a),
\end{equation}
in which we have identified $p_{_a}^2=\,-\frac{\hbar^2 G^2}{c^6}
a^{-s} \frac{d}{da}\Big(a^s\frac{d}{da}\Big)$, where the parameter
$s$ represents the ambiguity in the ordering of factors $a$ and
$p_a$ in the first term of the Hamiltonian and $\Psi(a)$ is the BH
wave function. By setting $R_{_s}=\frac{2GM}{c^2}$, one gets
\begin{equation}\label{wd2}
\frac{\hbar^2 G^2}{c^6} \frac{1}{a}
\Big(\frac{d^2}{d{a^2}}+\frac{2}{a}
\frac{d}{da}\Big)\Psi(a)\,=\,(a-R_{_s}) \Psi(a),
\end{equation}
in which we have set $s=2$.\footnote{Indeed, there are various
possibilities for ordering. Since the factor-ordering parameter will
not affect the semiclassical calculations in minisuperspace models,
one usually chooses a special value for it in a given model.} Using
the following transformations
\begin{eqnarray}\label{trans1}
\Psi(a)=\frac{1}{a} U(a)\qquad \mbox{and} \qquad y=a-R_{_s},
\end{eqnarray}
and bearing in mind that the energy of excited state associated with
the variable $a$ is not positive (this can be explained by the
physical argument that the total energy of the BH is included and
the ADM energy is equal to zero), the WDW equation takes the form
\begin{equation}\label{wd3}
\Big(-\frac{1}{2} {l_{_P}}^2 E_{_P} \frac{d^2}{d y^2}\,+\,\frac{
E_{_P}}{2{l_{_P}}^2}y^2\Big)U(y)\,=\,\frac{R_{_s}}{4 l_{_P}}U(y),
\end{equation}
where $E_{_s}=M c^2$ is the ADM energy of the Schwarzschild BH,
$l_{_P}=\sqrt{\frac{G \hbar}{c^3}}$ and $E_{_P}=\sqrt{\frac{c^5
\hbar}{G}}$ are the Planck's length and energy respectively.
Relation (\ref{wd3}) is the same as the Schr\"{o}dinger equation
for a one-dimensional harmonic oscillator with mass $m=M_{_P}=
\sqrt{\frac{\hbar c}{G}}$ and frequency $\hbar\omega=\sqrt{
\frac{3}{2\pi}} E_{_P}$. So, the quantized energy eigenvalues of
the Schwarzschild BH are given by
\begin{equation}\label{energybh}
\frac{R_{_s}(n)}{4 l_{_P}}\,E_{_s}=\,(n+\frac{1}{2})\,E_{_P}.
\end{equation}
The mass of the Schwarzschild BH is also quantized with respect
to the Planck mass as
\begin{equation}\label{massbh}
M^2(n)=(2n+1)\,M_{_P}^2\,,
\end{equation}
in agreement with the Bekenstein's proposal in the BHs
thermodynamics \cite{Bekenstein}. We use these results in our
forthcoming arguments.

\subsection{Hawking temperature and entropy}

In this section we discuss the Feynman's path integral method to
study thermodynamics of the Schwarzschild BH. In this method, one
deals with a semiclassical partition function instead of the
quantum one, and all of the quantum effects can be summarized in
a modified potential \cite{Feynman}. By the path integral method
applied to the harmonic oscillator, its potential energy and
classical partition function will change as follows \cite{Obregon}
\begin{equation}\label{mpotential}
U(x)=\frac{3 E_{_P}}{4\pi{l_{_P}}^2}\,\Big(x^2+\frac{\beta
{l_{_P}}^2 E_{_P}}{12}\Big),
\end{equation}
and
\begin{equation}\label{bhpf1}
Z=\sqrt{\frac{2\pi}{3}}\,\frac{\exp\Big[-\frac{\beta^2
E_{_P}^2}{16\pi}\Big]}{\beta E_{_P}}.
\end{equation}
One can obtain all the thermodynamical quantities of the
Schwarzschild BH through the partition function (\ref{bhpf1}). The
internal energy of the BH can be obtained by the standard definition
as
\begin{equation}\label{internalenergy}
\bar{E}\,=\,-\frac{\partial\ln[Z]}{\partial\beta}=\,Mc^2,
\end{equation}
where we have set explicitly $\bar{E}=Mc^2$. We note that in the
canonical ensemble the temperature is an external parameter.
Substituting the partition function (\ref{bhpf1}) into the relation
(\ref{internalenergy}), gives the relation between the temperature
and the mass of the BH
\begin{equation}\label{eqotemperature}
\beta^2-\frac{8\pi Mc^2}{E_{_P}^2}\beta+\frac{8\pi}{E_{_P}^2}=0.
\end{equation}
Solving the above equation for $\beta$, we arrive at the following
expression for the BH temperature in terms of its mass
\begin{equation}\label{temperature}
\beta=\,\beta_{_H}\,\Big(\,1\,-\,\frac{1}{
\beta_{_H}\,M c^2}\,\Big)\,,
\end{equation}
where $\beta_{_H}=\frac{8\pi M c^2}{E_{_P}^2}$ is the inverse of the
Hawking temperature and we have used the fact that $E_{_P}\ll\,
Mc^2$. The entropy of the BH can be obtained from the partition
function by the definition
\begin{eqnarray}\label{entropy1}
\frac{S}{k}=\,\ln[Z]\,-\,\beta\,\frac{\partial\ln[Z]}{\partial
\beta}\,.
\end{eqnarray}
Considering the partition function (\ref{bhpf1}) in this relation
and substituting the Hawking temperature, the entropy of the BH
becomes
\begin{eqnarray}\label{entropy2}
\frac{S}{k}=\,\frac{A_{_s}}{4l_{_P}^2}\Big(\,1\,-\,\frac{1}{8\pi}
\,\Big(\frac{E_{_P}}{M c^2}\Big)^2\,\Big)^2\,-\,\frac{1}{2}\,\ln
\bigg[\frac{A_{_s}}{4l_{_P}^2}\Big(\,1\,-\,\frac{1}{8\pi}\,\Big(
\frac{E_{_P}}{M c^2}\Big)^2\,\Big)^2\bigg]\,-\,\frac{1}{2}\ln[24]\,
+\,1\,,
\end{eqnarray}
where $A_{_s}=4\pi R_{_s}^2$ is the horizon area of the BH. Defining
the area of the BH in terms of the horizon area as
$A_{_{BH}}=A_{_s}\Big(\,1\,-\,\frac{1}{8\pi}\,\Big(\frac{ E_{_P}}{M
c^2}\Big)^2\,\Big)^2$, the above relation can be rewritten in the
following form
\begin{eqnarray}\label{entropy3}
\frac{S}{k}=\,\frac{A_{_{BH}}}{4l_{_P}^2}\,-\,\frac{1}{2}\,
\ln\bigg[\frac{A_{_{BH}}}{4l_{_P}^2}\bigg]\,+\,{\mathcal{O}}
\bigg[\Big(\frac{{M_{_P}}}{M}\Big)^2\bigg]\,.
\end{eqnarray}
As it is clear, the well-known logarithmic correction to the entropy
appears naturally in this method. This kind of correction term also
appears in the other approaches such as string theory
\cite{Mukherji}, canonical QG \cite{entropyCQG}, LQG
\cite{entropyloop}, GUP and noncommutative frameworks \cite{Kastrup,
Gour,Khosravi,entropyother,entropyGUP,jalalzadeh,Obregon2}.

\section{Hawking temperature and entropy in the polymer framework}

In the previous section we applied the Feynman's path integral
method to the semiclassical partition function in order to consider
the quantum effects in the thermodynamics of the Schwarzschild BH.
On the other hand, we obtained the modified semiclassical polymeric
partition function for a one-dimensional harmonic oscillator. So,
one can implement the partition function (\ref{pfolho}) to study the
thermodynamics of the Schwarzschild BH as a harmonic oscillator as
we have reviewed in the pervious section. It is also important to
note that our results are applicable only to Planck scale BH, since
the result (\ref{pfolho}) coincides with quantum ones in the high
temperature limit.

One can include the LQG effects in the path integral approach by
replacing the partition function (\ref{bhpf1}) with the polymeric
ones (\ref{pfolho}) as
\begin{equation}\label{ppf1}
Z^{\rm (poly)}=\,\sqrt{\frac{2\pi}{3}}\,\frac{\exp\Big[-\frac{
\beta^2E_{_P}^2}{16\pi}\Big]}{{\beta}E_{_P}}\,\Big(\,1\,+\,
\frac{\mu^2\,M_{_P}}{8\beta}\,+\,...\,\Big)\,,
\end{equation}
where we have substituted $m=M_{_P}$ for Schwarzschild BH when we
used the relation (\ref{pfolho}). Therefore, relation (\ref{ppf1})
shows the partition function of the quantum Schwarzschild BH in the
polymer framework and we may investigate the thermodynamics of the
BH using this relation. As usual, the definition
(\ref{internalenergy}) gives the relation between temperature and
mass of the Schwarzschild BH in polymer framework. Substituting the
polymeric partition function (\ref{ppf1}) into the relation
(\ref{internalenergy}), we obtain the following expression for the
BH temperature in the limit of $\mu\rightarrow\,0$,
\begin{equation}\label{PLeqotemperature}
\beta^3-\,\beta_{_H}\,\beta^2+\,\frac{\beta_{_H}}{Mc^2}\,\beta\,
=\,-\frac{\mu^2\beta_{_H}M_{_P}}{8Mc^2}\,,
\end{equation}
which upon solving this equation we find
\begin{equation}\label{PLtemperature}
\beta=\,\beta_{_H}\,\bigg(\,1\,-\,\frac{(1+\mu^2M_{_P}/8)}{
\beta_{_H}\,M c^2}\,+\,{\mathcal{O}}\bigg[\frac{{E_{_P}}^2}{
Mc^2}\bigg]\,\bigg).
\end{equation}
It is easy to see that in the limit of $\mu\rightarrow0$, the above
solution reduces to the non-deformed case (\ref{temperature}). The
next step is to calculate the entropy of the BH from partition
function (\ref{ppf1}) through the definition (\ref{entropy1}), which
yields
\begin{eqnarray}\label{PLEntropy1}
\frac{S^{\rm (poly)}}{k}=\,\frac{\beta^2{E_{_P}}^2}{16\pi}-
\frac{1}{2}\ln\bigg[\frac{\beta^2{E_{_P}}^2}{16\pi}\bigg]+
\ln\bigg[1+\frac{\mu^2 M_{_P}}{8\beta}\bigg]+\Big(1+\frac{
8\beta}{\mu^2 M_{_P}}\Big)^{-1}\,-\frac{1}{2}\ln[24]\,+\,1.
\end{eqnarray}
Substituting from the relation (\ref{PLtemperature}) into this
relation we find
\begin{eqnarray}\label{PLEntropy2}
\frac{S^{\rm (poly)}}{k}=\,\frac{A_{_s}}{4l_{_P}^2}\bigg(
\,1\,-\,\frac{(1+\mu^2 M_{_P}/8)}{8\pi}\,\Big(\frac{E_{_P}}{
Mc^2}\Big)^2\,\bigg)^2\,-\,\frac{1}{2}\,\ln\Bigg[\frac{
A_{_s}}{4l_{_P}^2}\bigg(\,1\,-\,\frac{(1+\mu^2 M_{_P}/8)
}{8\pi}\,\Big(\frac{E_{_P}}{M c^2}\Big)^2\,\bigg)^2\Bigg]\,
\nonumber\\+\,\frac{\mu^2{E_{_P}}^3}{32\pi Mc^4}\,\bigg(\,1
\,-\,\frac{(1+\mu^2 M_{_P}/8)}{8\pi}\,\Big(\frac{E_{_P}}{M
c^2}\Big)^2\,\bigg)^{-1}\,-\,\frac{1}{2}\ln[24]\,+\,1\,.
\end{eqnarray}
Defining the polymeric area of the BH in terms of the horizon area
as
\begin{eqnarray}\label{PLEntropy3}
A^{\rm (poly)}_{_{BH}}=\,A_{_s}\bigg(\,1\,-\,\frac{(1+
\mu^2 M_{_P}/8)}{8\pi}\,\Big(\frac{E_{_P}}{Mc^2}\Big)^2
\,\bigg)^2\,,
\end{eqnarray}
and notifying that $\frac{A^{\rm (poly)}_{_{BH}}}{A_{_s}}<1$, we
obtain the entropy of the quantum Schwarzschild BH in polymer
picture as
\begin{eqnarray}\label{PLEntropy4}
\frac{S^{\rm (poly)}}{k}=\,\frac{A^{\rm (poly)}_{_{BH}}}{
4l_{_P}^2}\,-\,\frac{1}{2}\,\ln\bigg[\frac{
A^{\rm (poly)}_{_{BH}}}{4l_{_P}^2}\bigg]\,+\,Mc^2\bigg(
1-\frac{A^{\rm (poly)}_{_{BH}}}{A_{_s}}\bigg)\,+\,{
\mathcal{O}}\Big[{A^{\rm (poly)}_{_{BH}}}^{-1}\Big]\,.
\end{eqnarray}
The above results are obtained in the high temperature limit since
our semiclassical consideration coincides with the full quantum
approach in this limit. As an important consequence, the
semiclassical, high-temperature result (\ref{PLEntropy4}) is
applicable only for Planck scale BHs such as the quantized
Schwarzschild BH that we studied previously in this paper. We see
that the relation (\ref{PLEntropy4}) is compatible with calculation
of entropy in other approaches such as GUP and noncommutative
geometry frameworks. Specially, the existence of logarithmic
correction term is a common feature in all these approaches (see
\cite{Mukherji,entropyCQG,
entropyloop,Kastrup,Gour,Khosravi,entropyother,entropyGUP,
jalalzadeh,Obregon2}). The sign of this correction is also
compatible with previous studies.

\section{Conclusions}
The polymer representation of quantum mechanics is an effective
model which has been investigated in a symmetric sector of LQG.
In this paper, we have developed statistical mechanics in the
polymer framework. The energy eigenvalues of microstates in
statistical mechanics are usually given by the Schr\"{o}dinger
equation in the standard Schr\"{o}dinger representation of the
quantum mechanics. However, solving the corresponding eigenvalue
problem in the polymer representation it is not an easy task due
to the nonlinear form of the Hamiltonian operator in this
picture. Therefore, we introduce a semiclassical method in which
one does not need to solve the eigenvalue problem, instead one
works with the Hamiltonian function and the density of states in
the classical phase space. Moreover, we have defined a particular
transformation in the polymeric phase space which transforms the
polymeric nonlinear Hamiltonian to the standard forms and
consequently all the polymeric effects summarized in the density
of states when one working with these new phase space variables.
Implementing this method, we obtain the polymeric partition
function and we have shown that our results coincide with full
quantum mechanical ones in the limit of high temperature.
Finally, we studied thermodynamics of quantum Schwarzschild BH
in the polymer framework by using of the Feynman path integral
approach in the semiclassical regime. We found the corrections
to the Hawking temperature and entropy due to the LQG effects.
These corrections are compatible with results obtained through
other approaches such as generalized uncertainty principle or
noncommutative geometry considerations.\\

{\bf Acknowledgement}\\
We would like to thank an anonymous referee for very insightful
comments.

\end{document}